\documentclass[13 pt]{article}
\usepackage{amssymb}
\newcommand{\one}[1]{\stackrel{1}{#1}}
\newcommand{\two}[1]{\stackrel{2}{#1}}
\newcommand{\betah}{\hat{\beta}}
\def\lp{ {\tilde p}}
\def\lq{ {\tilde q}}
\def\zbar{\overline z}
\def\hE{\hat {\cal E}}
\def\vpbar{\overline \varphi}

\def\hphi{\hat \phi}

\def\th{\theta}

\def\si{\sigma}

\def\vp{\varphi}

\newcommand{\pt}{\hat{p}}
\newcommand{\qt}{\hat{q}}
\newcommand{\beq}{\begin{equation}}
\newcommand{\eeq}{\end{equation}}
\newcommand{\bea}{\begin{eqnarray*}}
\newcommand{\eea}{\end{eqnarray*}}
\newcommand{\beqa}{\begin{eqnarray}}
\newcommand{\eeqa}{\end{eqnarray}}

\begin{document}
\newfont{\elevenmib}{cmmib10 scaled\magstep1}%

\newcommand{\preprint}{
            \begin{flushleft}
   \elevenmib Yukawa\, Institute\, Kyoto\\
            \end{flushleft}\vspace{-1.3cm}
            \begin{flushright}\normalsize  \sf
            YITP-02-39\\
           {\tt hep-th/0208005} \\ October 2002
            \end{flushright}}
\newcommand{\Title}[1]{{\baselineskip=26pt \begin{center}
            \Large   \bf #1 \\ \ \\ \end{center}}}
\hspace*{2.13cm}%
\hspace*{1cm}%
\newcommand{\Author}{\begin{center}\large
           P. Baseilhac\footnote{
pascal@yukawa.kyoto-u.ac.jp}\ \ \ \ \mbox{and}\ \ \ \ K.
Koizumi\footnote{kkoizumi@cc.kyoto-su.ac.jp}
\end{center}}
\newcommand{\Address}{{\baselineskip=18pt \begin{center}
           $^1$\it Yukawa Institute for Theoretical Physics\\
     Kyoto University, Kyoto 606-8502, Japan\vspace{0.4mm}\\
          $^2$\it Department of Physics,\\
           Kyoto Sangyo University, Kyoto 603-8555, Japan
      \end{center}}}
\baselineskip=13pt

\preprint
\bigskip

\Title{Sine-Gordon quantum field theory on the half-line\\  with
quantum boundary degrees of freedom}\Author

\vspace{- 0.1mm}
 \Address

\vskip 0.6cm

\centerline{\bf Abstract}\vspace{0.3mm} The sine-Gordon model on
the half-line with a dynamical boundary introduced by Delius and
one of the authors is considered at quantum level. Classical
boundary conditions associated with classical integrability are
shown to be preserved at quantum level too. Non-local conserved
charges are constructed explicitly in terms of the field and
boundary operators. We solve the intertwining equation associated
with a certain coideal subalgebra of $U_q(\widehat{sl_2})$
generated by these non-local charges. The corresponding solution
is shown to satisfy quantum boundary Yang-Baxter equations. Up to
an exact relation between the quantization length of the boundary
quantum mechanical system and the sine-Gordon coupling constant,
we conjecture the soliton/antisoliton reflection matrix and bound
states reflection matrices. The structure of the boundary state is
then considered, and shown to be divided in two sectors. Also,
depending on the sine-Gordon coupling constant a finite set of
boundary bound states are identified. Taking the analytic
continuation of the coupling, the corresponding boundary
sinh-Gordon model is briefly discussed. In particular, the
particle reflection factor enjoys weak-strong coupling duality.

\vspace{0.1cm} .
\vskip 0.8cm


\vskip -0.6cm

\section{Introduction}
Two-dimensional quantum field theories with boundary have
attracted attention for many years, as they play an important role
in the analysis of low dimensional statistical systems near
criticality or quantum gravity (open string approach). Solving
integrable theories restricted on the half-line for various types
of boundary is then of great interest. Among the known examples,
the sine-Gordon model with a non-dynamical boundary has been shown
to be integrable at classical level \cite{Skl87,Skl88} as well as
at quantum level \cite{Gho93,Gho94}. This has provided one of the
simplest examples for which exact results such as factorized
scattering theory, boundary spectrum,
etc...\cite{Fri95,Fen94,Sal95,Sko95,Lec95,Fat97,Dor99,Cor98,Cor99,Cor00,Dor00,Baj02,Mat}
have been derived.

Recently \cite{Bas01}, an integrable Hamiltonian which describes a
sine-Gordon model on the half-line coupled to a non-linear
oscillator at the boundary has been shown to be integrable at
classical level. Using a solution of the classical reflection
equations (also called classical boundary Yang-Baxter equations)
and following Sklyanin formalism we have obtained an infinite
number of mutually commuting classical conserved quantities,
provided some specific classical boundary conditions. The
existence of such integrals of motion is a sufficient condition
which ensures classical integrability of the model. Quantum
integrability had remained an open question. Although a solution
of the quantum reflection equations - quantum boundary Yang-Baxter
equations (qBYBE) - had been proposed for a certain representation
of the quantum $R$-matrix \cite{Kuz95,Kuz96}, there was no reason
to believe that this solution could describe the
soliton/antisoliton scattering process bouncing off the boundary.

Independently, the method of construction of non-local conserved
charges in integrable models in the bulk \cite{Ber91} has been
extended to non-dynamical boundary ones more recently. For
instance, the sine-Gordon model \cite{Mez98} and the $A^{(1)}_{r}$
affine Toda theory with imaginary coupling \cite{Del02} restricted
to the half-line have been studied. The existence of such
non-local conserved charges provides a useful tool to determine
the scattering properties of the theory. Indeed, the explicit form
of the $S$-matrix (soliton/antisoliton scattering) and the
$K$-matrix (soliton/antisoliton scattering on the boundary) is
encoded in the minimal solution of quantum Yang-Baxter equations
(qYBE)s and qBYBEs, respectively.

In this paper, we study at quantum level the model introduced in
\cite{Bas01}. In section 2 we will show that quantum boundary
conditions take a form similar to the classical boundary ones
proposed in \cite{Bas01}. Also, we will construct non-local
conserved charges corresponding to our dynamical case, which are
natural extensions of the known ones (non-dynamical)
\cite{Mez98,Del02}. In our case the non-local conserved charges
generate a certain coideal subalgebra of $U_q(\widehat{sl_2})$
mixed with the Heisenberg one, where $q$ is the deformation
parameter. Further, by specifying the representation for
$U_q(\widehat{sl_2})$ we solve the corresponding intertwining
equations. It provides a new solution $K_0(\theta)$ to the qBYBEs,
different from the one proposed in \cite{Kuz95,Kuz96,Bas01}.

In section 4, we use the minimal solution $K_0(\theta)$ of these
equations\,\footnote{These relations, in the dynamical case, are
no longer {\it linear} in terms of the whole set of operators.} to
construct the soliton/antisoliton reflection matrix
$K^{(\epsilon)}_{SG}(\theta)$  which describes the scattering of
the sine-Gordon soliton/antisoliton off the dynamical boundary. In
particular, taking the commutative limit of the Heisenberg
algebra, we check that the ``minimal'' part reduces to the one
obtained by Ghoshal-Zamolodchikov \cite{Gho93} and
DeVega-Gonzales-Ruiz \cite{DeV94}. In the generic noncommutative
case, we show that $K^{(\epsilon)}_{SG}(\theta)$ is a solution of
the qBYBEs associated with the sine-Gordon $S$-matrix. This shows
consistency of the nonperturbative analysis based on non-local
conserved charges approach. To determine uniquely
$K^{(\epsilon)}_{SG}(\theta)$ we impose the boundary unitarity
condition as well as the boundary cross-unitarity condition
proposed in \cite{Gho93}. Studying the singularities of the
proposed reflection matrix we identify a finite set of boundary
bound states. Then, we apply the bootstrap equation to construct
explicitly the reflection matrices associated with bound states
(breathers) scattering off the boundary.

As we are going to see, with the help of certain algebraic
relations between Heisenberg operators the boundary unitarity
condition is satisfied without specifying the boundary state
structure. However, the boundary cross-unitarity condition leads
to restrict its form in order to preserve certain ``physical''
principles.  We identify the corresponding constraints: choosing a
continuous representation for the boundary operators difference
equations are obtained. They restrict the boundary state in two
sectors, associated with ``even'' or ``odd'' breathers $B_n$.

Concluding remarks follow in the last section. There, following
Corrigan \cite{Cor98} we use the so-called breather-particle
identification to obtain the particle reflection factor in the
boundary sinh-Gordon model with quantum degrees of freedom at the
boundary. The result is found to be self-dual under the
weak-strong coupling duality transformation $b\leftrightarrow
2/b$.

\section{Classical sine-Gordon field theory with a dynamical boundary}
Let us first recapitulate some main results of \cite{Bas01}
without details. There, we considered a Hamiltonian describing the
interaction between a sine-Gordon field theory restricted to the
half-line and a non-linear oscillator living at the boundary. The
sine-Gordon part describes a relativistic 1+1 dimensional
self-interacting bosonic field $\phi(x,t)$ with mass $m$ and its
conjugate field $\pi(x,t)$. The Hamiltonian reads
\beqa\label{hsg}
  H_{SG}&=&\int_{-\infty}^0 dx\left(
  \frac{1}{2}\pi(x)^2+\frac{1}{2}(\partial_x\phi(x))^2-
  \frac{m^2}{\beta^2}\cos\beta\phi(x)\right) \\
  &&\ \ \ \ \ \ \ \  \ \ \ \  -\frac{2m}{\beta^2}\left(
  \cos(\beta \lp/\sqrt{2M_0 m})\ e^{-i\beta\phi(0)/2}+
  \cos(\beta\sqrt{M_0 m}\lq/2\sqrt{2})\,e^{i\beta\phi(0)/2}
  \right)\ .\nonumber
\eeqa
Expanding the oscillator part for small $\lq$ and $\lp$, these
degrees of freedom can be interpreted as position and momentum
variables, respectively. In this limit, it can be shown that the
effective mass of the oscillator depends on the value of the field
at the boundary and the parameter $M_0$.

At fixed time $t$ we introduced the Poisson bracket
\beqa \{{\cal O}_1,{\cal
O}_2\}=\int_{-\infty}^{0}\!dx\Big[\frac{\delta{\cal
O}_1}{\delta\pi(x)}
 \frac{\delta{\cal O}_2}{\delta\phi(x)}
- \frac{\delta{\cal O}_1}{\delta\phi(x)}
 \frac{\delta{\cal O}_2}{\delta\pi(x)}\Big] +  \frac{\partial {\cal O}_1}{\partial
\lp}\frac{\partial {\cal O}_2}{\partial \lq} - \frac{\partial
{\cal O}_1}{\partial \lq}\frac{\partial {\cal O}_2}{\partial
\lp}\label{def1} \eeqa
for any observable \ ${\cal O}_j$ . At constant time slices the
non-vanishing Poisson brackets in the sine-Gordon field theory and
for the boundary variables $\lp$, $\lq$ are, respectively
\beqa \{\pi(x),\phi(y)\}=\delta(x-y)\ ,\ \ \ \ \ \ \ \ \ \ \ \ \
\{\lp,\lq\}=1 \eeqa
from which it is straightforward to calculate the equations of
motion. However, in order to have the continuity of these
equations on the left half-line $]-\infty,0]$, the equations for
the field $\pi(x)$ leads to the classical boundary condition
\beqa\label{bc}
  \partial_x\phi(0)=-i\frac{m}{\beta}(
  \cos\pt\ e^{-i\beta\phi(0)/2}-\cos\qt\ e^{i\beta\phi(0)/2})\ .
\eeqa
where we have introduced the reduced parameters\ \
$\pt=\frac{\beta}{\sqrt{2M_0 m}}\lp$\ \ and \
$\qt=\frac{\beta\sqrt{M_0m}}{2\sqrt{2}}\lq$.

The model with Hamiltonian (\ref{hsg}) is integrable: it is
possible to construct an infinite number of higher spin integrals
of motion that are in involution with each other. For instance,
assuming the boundary condition above (\ref{bc}) the first
non-trivial integral of motion reads
\beqa\label{I3}
  I_3&=&\int_{-\infty}^0 \left(
    \frac{\beta^4}{16m^3}\left(\pi(x)^4+6\pi(x)^2(\partial_x\phi(x))^2+
    (\partial_x\phi(x))^4\right)
    -\frac{\beta^2}{m^3}\left((\partial_x\pi(x))^2+(\partial_x^2\phi(x))^2\right)
    \right.\nonumber\\
    && \left.
    -\frac{\beta^2}{4m}\left(\pi(x)^2+5(\partial_x\phi(x))^2\right)\cos\beta\phi(x)
    +\frac{m}{8}\left(\cos 2\beta\phi(x)\right)
    \right)dx + I_3^{boundary}
\eeqa
with
\bea I_3^{boundary}
&=&e^{3i\beta\phi(0)/2}\left(\frac{1}{2}\cos\qt
    +\frac{1}{6}\cos^3\qt\right)-e^{i\beta\phi(0)/2}\left(\frac{3}{2}\cos\pt
    -\frac{1}{2}\cos^2\qt\cos\pt
    +\frac{\beta^2}{2m^2}\pi(0)^2\cos\qt\right)\nonumber\\
    &+&e^{-3i\beta\phi(0)/2}\left(\frac{1}{2}\cos\pt
    +\frac{1}{6}\cos^3\pt\right)-e^{-i\beta\phi(0)/2}\left(\frac{3}{2}\cos\qt
    -\frac{1}{2}\cos^2\pt\cos\qt
    +\frac{\beta^2}{2m^2}\pi(0)^2\cos\pt\right)\nonumber\\
    &-&i\frac{2\beta}{m}\sin\qt\,\sin\pt\,.\nonumber
\eea
In general, this Hamiltonian is not real.  Such kind of situation
is typical in higher rank affine Toda theories with imaginary
coupling for which the Hamiltonian is also non-hermitian. In these
theories, the classical soliton solutions are found to be complex.
Nevertheless, the energy of these configurations is real
\cite{Hol92}.

\section{Quantum sine-Gordon field theory with a quantum boundary}
Let us suppose that there exists a well-defined action describing
the sine-Gordon (SG) field theory restricted on the half-line
coupled with a quantum mechanical system at the boundary. To
characterize the dynamics of the boundary we introduce, let say,
two boundary operators ${\cal E}_{\pm}(y)$. We propose the
following action
\beqa {\cal A}=\int_{-\infty}^{\infty}dy\int_{-\infty}^{0}dx
\Big(\frac{1}{8\pi}(\partial_\nu\phi)^2-2\mu\cos(\betah\phi)\Big)
+ \mu^{1/2}\int_{-\infty}^{\infty}dy\ \Phi_{pert}^{B}(y)\ + \
{\cal A}_{boundary} \label{action} \eeqa
where the interaction between the field and the dimensionless
boundary quantum operators reads
\bea \Phi_{pert}^{B}(y) \ =\ {\cal E}_-(y) e^{i\betah\phi(0,y)/2}
+ {\cal E}_+(y) e^{-i\betah\phi(0,y)/2}\  \eea
and ${\cal A}_{boundary}$ is the kinetic part associated with the
boundary operators. Here we used the notation
$\betah=\beta/\sqrt{4\pi}$. In the deep UV, this model can be
considered as a relevant perturbation of a conformal field theory
(for instance the free Gaussian field) on the semi-infinite plane
with certain boundary conditions at $x=0$. In the
limit\,\footnote{As we will see later, integrability of the QFT
(\ref{action}) requires that one can not turn on the boundary
perturbation independently of $\mu$.} $\mu=0$ Neumann boundary
conditions $\partial_x\phi|_{x=0}=0$ preserve integrability. If
one turns on the boundary perturbation ($\mu\neq 0$), following
the arguments of \cite{Gho93} we expect that a quantum analogue of
the classical integral of motion (\ref{I3}) can be constructed
explicitly. Obviously, the boundary operators $\hE_\pm(y)$ will
have to satisfy certain algebraic relations in order to preserve
integrability at quantum level {\it too}. Even if we do not
construct explicitly such local quantum conserved charges here,
analysis of following sections will support integrability of the
model.

There are two alternative description\,\footnote{It seems
important to us to recall the analysis of \cite{Gho93}.} of the
model (\ref{action}). On one hand, one can choose the
$y$-direction to be the ``time'' in which case the Hamiltonian
$H_B$ contains the boundary contribution, and the Hilbert space
${\cal H}_B$ is identified with the half-line $y=Const.$. Then,
correlation functions are calculated over the ground state of
$H_B$, denoted $|0\rangle_B$ below.

On the other hand, one can choose the infinite line $x=Const.$ as
the ``equal time section''. The Hamiltonian in this case is the
same as in the bulk theory on the full line with Hilbert space
${\cal H}$. The boundary at $x=0$ plays the role of initial
condition, and all its information is encoded in the boundary
state $|B\rangle\in {\cal H}$. Correlation functions are written
as $\langle 0|...|B\rangle$ where $|0\rangle\in {\cal H}$ is a
(degenerate) ground state of ${\cal H}$. Notice that the existence
of local integrals of motion ${\cal I}^{(s)},{\overline{\cal
I}}^{(s)}$ for certain values of $s$ implies
\beqa ({\cal I}^{(s)}-{\overline{\cal I}}^{(s)})|B\rangle = 0\
.\label{cond1}\eeqa

In the perturbed boundary conformal field theory (BCFT) approach,
the free bosonic fundamental field restricted to the half-line can
be written in terms of its holomorphic/anti-holomorphic components
\bea \phi(x,y)=\vp(z)+\vpbar(\zbar)\ .\eea
If we denote the expectation value in the BCFT with Neumann
boundary conditions $\langle ... \rangle_0$, then the holomorphic
components are normalized such that
\bea \langle{\varphi}(z){{\varphi}}(w)\rangle_0=-2\ln(z-w),\quad
\langle{{\bar\varphi}}({\bar z}){{\bar\varphi}}({\bar
w})\rangle_0=-2\ln({\bar z}-{\bar w}),\quad
\langle{\varphi}(z){{\bar\varphi}}({\bar
w})\rangle_0=-2\ln(z-{\bar w})\ .\label{correl} \eea
To find the boundary conditions at quantum level, we can consider
the expectation value of the local field $\partial_x\phi(0,y)$
with any other local field in first order of boundary conformal
perturbation theory $\mu\rightarrow 0$. Due to the presence of the
boundary operators ${\cal E}_{\pm}(y)$, we write the ground state\
\ $|0\rangle_B$\ as
\beqa |0\rangle_B\ \ =\ \ |vac\rangle_B\otimes|0\rangle_{BCFT}\ \
\ +\ \ \ {\cal O}(\mu) \ ,\label{vac}\eeqa
where \ \ $|0\rangle_{BCFT}\in {\cal H}_{BCFT}$\ \ and \ \
$|vac\rangle_B$ \ denotes the effective boundary ground state
which satisfies the Schr\"{o}dinger equation
\beqa H({\cal E}_+,{\cal E}_-)|vac\rangle_B =
E^{eff}_{0,B}|vac\rangle_B \ \eeqa
with effective ground state\,\footnote{Taking the perturbative
limit $\betah\rightarrow 0$ and using an explicit realization of
the boundary operators in terms of Heisenberg operators (see
further sections), one can show that the ``pure'' boundary wave
function satisfies certain difference equations.} energy
$E_{0,B}^{eff}$. Here, $H({\cal E}_+,{\cal E}_-)$ is an effective
boundary Hamiltonian.
In this approximation, following the analysis of \cite{Del02} with
eqs. (\ref{correl}) it is straightforward to show that, in first
order of boundary conformal perturbation theory, the quantum
boundary condition takes a form similar to the classical one.
Furthermore, using scaling arguments\,\footnote{Dimensions of both
side must be equal. Due to the form of the perturbing operator one
finds that the only term that can appear on the right hand side
must be linear in the parameter $\mu$. All other terms must
vanishes.} it is possible to show that this condition is satisfied
at {\it all} order in perturbation theory. It reads
\beqa\label{bcq}
  _B\langle0|\partial_x\phi(0,y)...|0\rangle_B=i2\pi\betah\mu^{1/2}\ _B\langle0| ({\cal
  E}_-(y)
  e^{i\betah\phi(0,y)/2}- {\cal E}_+(y) e^{-i\betah\phi(0,y)/2})...|0\rangle_B\ . \eeqa
The method of constructing non-local conserved charges in the
model (\ref{action}) follows the line presented in \cite{Del02}.
So, we refer the reader to this paper for more details. Here, the
main difference is that ${\cal E}_\pm(y)$ are operators. If we
denote the ``bulk'' non-local charges \cite{Ber91}
\bea
Q_{\pm}=\frac{1}{4\pi}\int^{\infty}_{-\infty}dx(J_{\pm}-H_{\pm})\
,\qquad \qquad {\bar
Q}_{\pm}=\frac{1}{4\pi}\int^{\infty}_{-\infty}dx({\bar
J}_{\pm}-{\bar H}_{\pm})\ , \eea
they can be expressed in terms of holomorphic/anti-holomorphic
part of vertex operators as\,\footnote{On the contrary to
\cite{Del02}, here we denote the fundamental field one the {\it
whole} line $\tilde{\phi}(x,y)$ and the one restricted to the
half-line $\phi(x,y)$. Then, the chiral components are related in
the following way: ${\varphi}(x,y)={\tilde{\varphi}}(x,y)+{
\tilde{\vpbar}}(-x,y),\quad
{\vpbar}(x,y)={\tilde{\vpbar}}(x,y)+{\tilde\varphi}(-x,y)$.}
\bea &&
J_{\pm}=:\exp({\pm\frac{2i}{\betah}{\tilde\varphi}}):\qquad
\mbox{and}\qquad
H_{\pm}=-4\pi\mu\frac{\betah^2}{\betah^2-2}:\exp\left(\pm
i\left(\frac{2}{\betah}-\betah\right){\tilde\varphi}\mp
i\betah{{\tilde{\vpbar}}}\right):\ ,\nonumber\\
&& {\bar J}_{\pm}=:\exp({\mp
\frac{2i}{\betah}{{\tilde{\vpbar}}}}): \qquad \mbox{and} \qquad
{\bar H}_{\pm}=-4\pi\mu\frac{\betah^2}{\betah^2-2}:\exp\left(\mp
i\left(\frac{2}{\betah}-\betah\right){{\tilde{\vpbar}}}\mp
i\betah{\tilde\varphi}\right):\ . \eea
Together with the ``bulk'' topological charge
$T_{bulk}=\frac{\betah}{2\pi}\int^{\infty}_{-\infty}dx\ \partial_x
{\tilde\phi}$ \
they generate the quantum enveloping algebra $U_q(\widehat{
sl_2})$. In the theory on the half-line, these charges are no
longer conserved. Instead, using (\ref{vac}) and the method of
\cite{Mez98,Del02} it is straightforward to show that the two
combinations
\beqa {\hat Q}_{\pm}=Q_{\pm}+{\bar Q}_{\mp}+{\hat{\cal
E}}_{\pm}(y)q^{\mp T}\ \label{nonloc}\qquad
\qquad\mbox{with}\qquad \qquad
T=\frac{\betah}{2\pi}\int^{0}_{-\infty}dx\ \partial_x {\phi}\eeqa
and \vspace{-0.5cm}
\beqa {\hat{\cal
E}}_{\pm}(y)=\mu^{1/2}\frac{\betah^2}{1-\betah^2}{\cal E}_\pm(y)
\label{operat}\eeqa
are conserved to all orders in boundary conformal perturbation
theory framework. Here, the deformation parameter is defined by
\beqa q\equiv\exp(-2i\pi/\betah^2)\ .\label{defor}\eeqa

As the sine-Gordon model possesses a single two dimensional
soliton multiplet denoted by $|\psi_{\pm}(\theta)\rangle$, we are
now interested in two dimensional representations of
$U_q(\widehat{ sl_2})$ on asymptotic soliton states, denoted
$\pi_\theta$ below. Here, $\theta$ denotes the rapidity of the
soliton/antisoliton. To determine the meaner in which non-local
conserved charges (\ref{nonloc}) are represented on asymptotic
single-soliton states, we use the notation of \cite{Ber91}.
Furthermore, in the asymptotic limit $y\rightarrow\pm\infty$ on
the boundary we will assume
\beqa {\hat{\cal E}}_{+}(y=\pm\infty)\equiv{\hat{\cal
E}}_{+}\qquad \qquad \mbox{and}\qquad \qquad {\hat{\cal
E}}_{-}(y=\pm\infty)\equiv{\hat{\cal E}}_{-}\ .\label{opcond}\eeqa
In the first Hamiltonian picture  $y$ is identified with the time
axis. In other words, the condition (\ref{opcond}) means that
boundary operators lead to the same operators in the far past and
future. We must keep in mind that $\hE_{\pm}$ are operators
satisfying \ $[\hE_{\pm},g]=0$ \ and \ $\pi_\theta(\hE_\pm
g)=\hE_\pm \pi_\theta(g)$, \ $\forall g\in U_q(\widehat{sl_2})$.
Thus we have
\beqa \pi_\theta({\hat Q}_{\pm})^+_+={\hat {\cal E}}_\pm q^{\mp1}\
, \qquad \pi_\theta({\hat Q}_{\pm})^+_-=c e^{\pm\lambda\theta}\ ,
\qquad \pi_\theta({\hat Q}_{\pm})^-_+=c e^{\mp\lambda\theta} \
,\qquad \pi_\theta({\hat Q}_{\pm})^-_-= {\hat {\cal E}}_\pm q^{\pm
1}\ \nonumber \eeqa
with
\beqa \lambda=2/\betah^2-1\qquad \ \ \ \ \mbox{and}\ \ \ \ \qquad
c^2=i2\mu(q^2-1)/\lambda^2 \ .\eeqa
For the remaining symmetry algebra of the QFT (\ref{action})
generated by the non-local conserved charges given above
(\ref{nonloc}) we are looking for a solution to the intertwining
equation
\beqa {K_0}^\delta_\nu(\theta){\pi_\theta({\hat
Q}_{\pm})}^\nu_\zeta={\pi_{-\theta}({\hat
Q}_{\pm})}^\delta_\nu{K_0}^\nu_\zeta(\theta) \label{inter}\eeqa
where indices $\{\delta,\nu,\zeta\}\in\{\pm\}$ refer to two
dimensional representations of $U_q(\widehat{ sl_2})$ on
asymptotic soliton states $|\psi_{\pm}(\theta)\rangle$. Then,
using this representation the solution $K_0(\theta)$ is written as
a $2\times 2$ matrix with entries expressed in terms of the
boundary operators. In the non-dynamical case, the entries are
just analytic functions of $\theta$ as ${\cal E}_\pm$ are
$c-$numbers. For further convenience, let us define
\beqa {K_0}^+_+(\theta)=A(\theta)\ ,\qquad \qquad
{K_0}^+_-(\theta)=B(\theta)\ ,\nonumber \\
{K_0}^-_+(\theta)=D(\theta)\ ,\qquad \qquad
{K_0}^-_-(\theta)=E(\theta)\ .\label{def}\eeqa
After some calculations, we find that the entries of the minimal
solution ${K_0}(\theta)$ of the intertwining property
(\ref{inter}) takes the following form:
\beqa
&&A(\theta)=\big(q^{-1}e^{\lambda\theta}\hE_+-qe^{-\lambda\theta}\hE_-\big)(q-q^{-1})/2c\ ,\nonumber \\
&&E(\theta)=\big(q^{-1}e^{\lambda\theta}\hE_--qe^{-\lambda\theta}\hE_+\big)(q-q^{-1})/2c\ ,\nonumber \\
&&B(\theta)=\Big(-c^2q^{-1}e^{2\lambda\theta}-c^2qe^{-2\lambda\theta}+\frac{q-q^{-1}}{q+q^{-1}}(q^{-1}\hE_-\hE_+
 - q\hE_+\hE_-)\Big)/2c^2\ ,\nonumber\\
&&D(\theta)=\Big(-c^2q^{-1}e^{2\lambda\theta}-c^2q
e^{-2\lambda\theta}+\frac{q-q^{-1}}{q+q^{-1}}(-q\hE_-\hE_+
 + q^{-1}\hE_+\hE_-)\Big)/2c^2\ ,\label{entrees}
\eeqa
if the boundary operators satisfy certain commutation relations
with the entries $B(\theta)$ and $D(\theta)$. Among these, we
obtain the commutation relation\,\footnote{For generic values of
the coupling and explicit boundary operators dependence in the
boundary scattering description, this condition is necessary.
Also, we do not assume specific values of the coupling here (e.g.
free fermion point, reflectionless point), for which new solutions
may be constructed. }
\beqa \big[\hE_+\hE_-,\hE_-\hE_+\big]= c^2(q+q^{-1})^2\big(\hE^2_+
-\hE^2_-\big) \ .\label{commut} \eeqa
Here, we propose a realization of the boundary operators at
$y=\pm\infty$ in terms of operators $\{\pt,\qt\}$ which belong to
the Heisenberg algebra \ $\big[\pt,\qt \big]\sim\alpha$\ \ in
order to have (\ref{commut}) satisfied. Due to (\ref{opcond}) we
denote \ $\pt(\pm\infty)\equiv\pt$\ and \
$\qt(\pm\infty)\equiv\qt$. Indeed, one can check that
setting\,\footnote{Notice that this solution also works,
obviously, if one changes $\alpha\rightarrow -\alpha$.}
\beqa&&\hE_+=\pm 2e_{UV}\cosh\pt \ \ \ \ \ \mbox{and}\ \ \ \ \
\hE_-=\pm 2e_{UV}\cosh\qt \ \ \ \ \ \mbox{where}\ \ \ \ \pm
q=\exp(\alpha/2)\label{defope0}\eeqa
with the normalization $e_{UV}^2=-c^2/(q-q^{-1})^2$, all the
equations (\ref{inter}) can be satisfied for a certain relation
between the SG coupling constant and the boundary quantization
length $\alpha$. However, according to eqs. (\ref{defope0}) there
is an ambiguity in the definition of the sign of the boundary
perturbation.
Depending on each sign, the boundary quantization length is fixed
to
\beqa \big[\pt,\qt \big]\!\!&=&\!\!\alpha\ mod\ (4i \pi) \ \ \ \ \
\mbox{with}\ \ \  \ \ \alpha=i4\pi\frac{(\betah^2-1)}{\betah^2}
\quad (+) \ \ \ \ \ \ \mbox{or}\ \ \ \ \ \ \alpha=i
2\pi\frac{(\betah^2-2)}{\betah^2} \quad (-)\ .\label{relat} \eeqa

In the model (\ref{action}) with (\ref{operat}) and
(\ref{defope0}), $e_{UV}$ is fixed by the ``bulk'' non-local
conserved charges algebraic structure which  gives
\beqa e^2_{UV}= \frac{i2(1-q^2)}{\lambda^2(q-q^{-1})^2}\ \mu\eeqa
To relate the UV parameter $e_{UV}$ and the IR soliton mass
parameter $M$ we can use their exact relation calculated in
\cite{Zam95}:
\beqa \mu=\frac{\Gamma(\betah^2/2)}{\pi\Gamma((2-\betah^2)/2)}
\Big[\frac{M\sqrt{\pi}\Gamma(1/2+1/2\lambda)}{2\Gamma(1/2\lambda)}\Big]^{2-\betah^2}\
. \label{massmu} \eeqa
It follows that the strength of the boundary perturbation is fixed
by the bulk soliton mass $M$. This is consistent with the
classical model (\ref{hsg}) in the sense that the only free
parameter is the mass $m$\,\footnote{Whatever the mass of the
effective oscillator is, the relevant quantity is its frequency
$\omega=m/2$ \cite{Bas01}. Using the particle-breather
identification, one can relate $m$ associated with the lightest
breather with the mass $m$ of the fundamental particle.}. In other
words, choosing an arbitrary boundary mass would destroy
integrability. Physically, this phenomena is not so surprising: to
preserve integrability the effects brought by the SG model and the
non-linear oscillator (energy transfer, for instance) have to
compensate each other, which is characterized by the exact
relation between the quantization length $\alpha$ and the SG
coupling constant in (\ref{relat}).

At specific values of the SG coupling constant $\betah^2=2/(n+1)$\
\ with $n\in\mathbb{N}$ \ the operators $\hE_\pm$ commute, and we
have $q^2=1$. Then, there is no need to assume some special
relation involving $c$ and the boundary parameters like
(\ref{commut}), and $\hE_\pm$ remain free. Consequently,
$K_0(\theta)$ is reduced to a $2\times 2$ matrix where $\hE_\pm$
are $c-$numbers. Expanding $K_0(\theta)$ in powers of $(q^2-1)$ we
obtain
\bea K_0(\theta)\sim 2\cosh(2\lambda\theta)\mathbb{P}/(q-q^{-1})+
iK^{min}_{non-dyn}(\theta)\ ,\eea
where $\mathbb{P}$ is the $2\times 2$ permutation matrix. The
first (singular) term is a trivial solution of the intertwining
property (\ref{inter}) whereas the second term is the minimal
reflection matrix obtained by Ghoshal-Zamolodchikov and
DeVega-Gonzales-Ruiz \cite{Gho93,DeV94}.

\section{Boundary Yang-Baxter equations and factorized scattering theory}
For the coupling constant $\betah^2<2$, the bulk sine-Gordon model
in 1+1 Minkowski space-time is massive and integrable. The
particle spectrum consists of a soliton/antisoliton pair
$(\psi_+(\theta),\psi_-(\theta))$ with mass $M$ and neutral
particles, called ``breathers'', $B_n(\theta)$\ \
$n=1,2,...,<\lambda$. As usual, $E=M\cosh\theta$ and
$P=M\sinh\theta$ the energy and momentum, respectively, of the
soliton/antisoliton. In this model, ${\cal H}$ is the Fock space
of multiparticle states. A general $N$-particles state is
generated by the ``particle creation operators''
$A_{a_i}(\theta_i)$
\beqa |A_{a_1}(\theta_1)...A_{a_N}(\theta_N)\rangle=
A_{a_1}(\theta_1)...A_{a_N}(\theta_N)|0\rangle\label{asymst}\eeqa
where $a_i$ characterizes the type of particle. The commutation
relations between these operators are determined by the $S$-matrix
elements, which describe the factorized scattering theory
\cite{Zam79}. Integrability imposes strong constraints on the
system which implies that the $S$-matrix has to satisfy the
quantum Yang-Baxter equations. For instance, the SG
soliton/antisoliton scattering $S$-matrix can be written as
\cite{Zam79}
\beqa S_{SG}(\theta)=R(\theta)\rho(-i\theta)/i\
,\label{sgmat}\eeqa
where we introduced the four dimensional representation of the
trigonometric solution of the quantum Yang-Baxter equation written
in terms of the standard Pauli matrices $\sigma_k$, \ $k=1,2,3$:
\beqa R(\th)=\frac{a(\th)}{2}\big(I\otimes I + \si_{3}
\otimes\si_{3}\big) + \frac{b(\th)}{2} \big(I\otimes I - \si_{3}
\otimes\si_{3}\big)+ \frac{c(\th)}{2} \big(\si_1\otimes \si_1 +
\si_2 \otimes\si_2\big)\  \eeqa
with
\bea a(\th)=\sinh(i\lambda\pi-\lambda\th)\ \quad,\quad\
b(\th)=\sinh(\lambda\th)\ \quad,\quad\ c(\th)=\sinh(i\lambda\pi)\
. \eea
The amplitudes $b(\theta)$ and $c(\theta)$ possess simple poles in
the physical strip\  $0< \theta < i\pi$\  located at
$\theta_n=i\pi-in\pi/\lambda$\ for \ $n=1,2,...<\lambda$. They are
interpreted as the neutral bound state (breather) $B_n$. The
factor $\rho(u)$ in (\ref{sgmat}) ensures unitarity and crossing
symmetry of the $S$-matrix. It reads
\beqa \rho(u)=-\frac{1}{\pi}\Gamma(\lambda)\Gamma(1-\lambda
u/\pi)\Gamma(1-\lambda+\lambda
u/\pi)\prod_{l=1}^{\infty}\frac{F_l(u)F_l(\pi-u)}{F_l(0)F_l(\pi)}\
, \eeqa
with
\bea F_l(u)=\frac{\Gamma(2l\lambda-\lambda
u/\pi)\Gamma(1+2l\lambda-\lambda
u/\pi)}{\Gamma((2l+1)\lambda-\lambda
u/\pi)\Gamma(1+(2l-1)\lambda-\lambda u/\pi)}\ .\eea
Using the bootstrap equation, the amplitudes associated with
(anti)soliton-breather $\psi_\pm B_n$ and breather-breather
$B_nB_m$ scattering have been calculated in \cite{Zam79}.

\subsection{Soliton reflection matrix for the dynamical boundary}
Taking the first Hamiltonian picture, we can consider $y=it$ in
the QFT (\ref{action}). If one uses the approach of \cite{Gho93},
asymptotic states are now generated using $A_a(\theta)$ acting on
the ground state $|0\rangle_B\in{\cal H}_B$. For boundary
integrable field theories, the action of the creation operator on
the boundary is characterized by the reflection matrix. It
describes the scattering of the soliton/antisoliton on the
boundary and is constrained by the so-called reflection equations,
i.e. qBYBEs for a certain choice of representation. Depending of
the quantization condition in (\ref{relat}) which determines the
sign $\epsilon\in\{\pm\}$ of the boundary operators
(\ref{defope0}) - the sine-Gordon reflection matrix
$K^{(\epsilon)}_{SG}(\theta)$ must satisfy the qBYBEs
\beqa\label{qKK} R(\th-\th'){\one
{K^{(\epsilon)}_{SG}}}(\th)R(\th+\th'){\two
{K^{(\epsilon)}_{SG}}}(\th') ={\two
{K^{(\epsilon)}_{SG}}}(\th')R(\th+\th'){\one
{K^{(\epsilon)}_{SG}}}(\th)R(\th-\th')\  \eeqa
where we used the  notations ${\one K}=K\otimes I$,
 ${\two K}=I \otimes K$. Recalling the definition of the entries (\ref{def}) we have
fourteen functional equations:
\bea
 (i)&&a_-c_+
 \left(B D' -B' D \right)+a_-a_+
 [A ,A']=0\ ,\\
(ii) &&b_-b_+
 [A, E']+
 c_-c_+
 [E ,E']+\ c_-a_+
 \big(D B' -D' B \big)=0\ ,\\
(iii)&&c_-b_+
 \big(E A'-E'A \big)+ b_-c_+
 \big(A A'-E'E\big)+\ b_-a_+
 [B,D']=0\ ,\\
(iv)&&b_-b_+AD'  +
 c_-c_+ED'+\ c_-a_+DA'
 -\ a_-a_+D' A -\ a_-c_+E'D=0\ ,\\
 (v)&&b_-b_+B'A+\ c_-c_+ B'E+\ c_-a_+A'B-\ a_-a_+A B'-\ a_-c_+BE'= 0\ ,\\
(vi)&&b_-b_+D'E+\ c_-c_+D'A+\ c_-a_+E'D-\ a_-a_+ED'-\ a_-c_+DA'=0\ ,\\
(vii)&&b_-b_+EB'+\ c_-c_+AB'+\ c_-a_+BE'-\ a_-a_+B'E-\ a_-c_+A'B= 0\ ,\\
(viii)&&b_-a_+BE' +
 \ c_-b_+EB'+\ b_-c_+AB'
 -\ a_-b_+E'B=0\ ,\\
(ix)&&b_-a_+A'B+\ c_-b_+B'A+\ b_-c_+B'E-\ a_-b_+BA'=0\ ,\\
(x)&&b_-a_+E'D+\ c_-b_+D'E+\ b_-c_+D'A-\ a_-b_+DE'=0\ ,\\
(xi)&&b_-a_+DA'+\ c_-b_+AD'+\ b_-c_+ED'-\ a_-b_+A'D=0\ , \eea
where we used the shorthand notations $a_-=a(\theta-\theta')$,
$a_+=a(\theta+\theta')$ and similarly for $b$ and $c$ as well as
$A=A(\theta)$ and $A'=A(\theta')$ and similarly for $B, D$ and
$E$. The remaining three equations are obtained from $(i)$,
$(ii)$, $(iii)$ through the substitutions $A\leftrightarrow E$ and
$B\leftrightarrow D$. Straightforward  calculations show that
$K^{(\epsilon)}_{SG}(\theta)$ takes a form similar to
 $K_0(\theta)$. Indeed, the reflection matrix in the dynamical boundary case can be written
\beqa
K^{(\epsilon)}_{SG}(\theta)=K^{(\epsilon)}_0(\theta)Y(\theta)\
,\label{ksg} \eeqa
where the function $Y(\theta)$ has to be determined using some
physical assumptions (see below). In the two dimensional
representation of $U_q(\widehat{sl_2})$, the matrix
$K^{(\epsilon)}_0$ is obtained from (\ref{entrees}) with the
following substitutions\,\footnote{Notice that for specific values
of the SG coupling constant such that $q$ is a root of unity, it
is easy to construct a finite dimensional representation for the
operators $A$, $B$, $D$, $E$ in the reflection matrix.}:
\beqa\hE_+\rightarrow \epsilon\cosh\pt \ ,\ \ \ \ \ \
\hE_-\rightarrow \epsilon\cosh\qt\ , \ \ \ \ \ q\rightarrow
e^{i\pi\lambda} \ \ \ \ \ \mbox{and}\  \ \ \ c\rightarrow
\sin(\pi\lambda)\ .\label{defope}\eeqa
Here, the quantization condition is given in (\ref{relat}). In
\cite{Gho93}, Ghoshal and Zamolodchikov proposed the use of the
``boundary unitarity'' and ``boundary cross-unitarity'' conditions
to determine the overall factor $Y(\theta)$ associated with a
non-dynamical boundary. Then, let us first consider the boundary
unitarity condition. Independently of any representation for the
boundary ground state $|0\rangle_B$, in our case it is natural to
consider the combination of operators
\beqa\label{unitarity0}
{K^{(\epsilon)}_{SG}}_c^a(\theta){K^{(\epsilon)}_{SG}}_b^c(-\theta)=
\delta_b^a\mathbb{I}\qquad\ \ \mbox{with}\ \ \qquad
\{a,b,c\}\in\{\pm\}\eeqa
as a generalization to the dynamical boundary case of the
condition associated with the non-dynamical one. Here, the
operator $\mathbb{I}$ denotes the identity which acts trivially on
the boundary ground state. In terms of the entries defined in
(\ref{def}) these relations becomes
\beqa
&&A(\theta)A(-\theta)+B(\theta)D(-\theta)=\mathbb{I}/Y(\theta)Y(-\theta)\ ,\nonumber\\
&&D(\theta)B(-\theta)+E(\theta)E(-\theta)=\mathbb{I}/Y(\theta)Y(-\theta)\ ,\nonumber\\
&&A(\theta)B(-\theta)+B(\theta)E(-\theta)=0\ ,\nonumber\\
&&D(\theta)A(-\theta)+E(\theta)D(-\theta)=0\ .\nonumber \eeqa
As the entries are noncommuting between each others, in general
one would expect the product (\ref{unitarity0}) to be ill-defined.
However, due to the specific form of the boundary operators one
can show that changing $\theta\rightarrow -\theta$ does not affect
the relations above. Also, the first two relations give the same
result. Then, using algebraic relations like
\beqa &&(q^2+q^{-2})\hE_+\hE_-\hE_+ - \hE_+^2\hE_- - \hE_-\hE_+^2
= -c^2(q+q^{-1})^2\hE_-\ ,\nonumber \\
&& (q^2+q^{-2})\hE_-\hE_+\hE_- - \hE_+\hE_-^2 - \hE_-^2\hE_+ =
-c^2(q+q^{-1})^2\hE_+\nonumber \eeqa
after some calculations we arrive at the condition
\beqa
Y(\theta)Y(-\theta)=\big[\sinh^2(2\lambda\theta)-\sinh^2(i\pi\lambda)\big]^{-1}\
\label{unitarity} . \eeqa
Compared to the non-dynamical boundary sine-Gordon, the situation
here is rather simple: there are no free parameters left in the
unitarity condition (\ref{unitarity}) appart from the coupling
constant $\betah^2$ which appears through $\lambda$.\vspace{0.2cm}

Let us now consider the boundary cross-unitarity condition. In the
alternative Hamiltonian description, $x$ is now interpreted as
``time'' and the space of states is the same as in the bulk
theory, i.e. ${\cal H}$. The ``initial'' (boundary) condition at
$x=0$ is encoded in $|B\rangle$. If the theory is integrable
(\ref{cond1}) must be satisfied. It follows that $|B\rangle$ is a
superposition of asymptotic states of the bulk theory constituted
by pairs of particles of equal mass but opposite rapidities
\cite{Gho93}. Thus, we define the boundary state as
\beqa |B\rangle = {\cal N}\Big(\ |{\small\textsc{B}}_0\rangle\ +\
\sum_{\{n\}}g^n B_n(0)|{\small\textsc{B}}_n\rangle\ +\
\frac{1}{2}\int_{-\infty}^{+\infty}d\theta
{K^{(\epsilon)}}_{\overline
a}^b(i\frac{\pi}{2}-\theta)A_{a}(\theta)A_{b}(-\theta)|{\small
\textsc{B}}_{ab}\rangle +...\Big)\ \ \label{stategen}\ .\eeqa
Here ${\cal N}$ is a normalization coefficient and the states
$|{\small\textsc{B}}_{a_1...a_N}\rangle\in{\cal
V}_{{\small\textsc{B}}}$ where ${\cal V}_{{\small\textsc{B}}}$ is
a representation for boundary operators. The second term
corresponds to the contribution from the zero-momentum particles,
namely the breathers. The coefficients $g^n$ indicate their
contribution. Considering in particular the soliton scattering,
arguments of \cite{Gho93} can be applied and we assume the
boundary cross-unitarity condition
\beqa {K^{(\epsilon)}_{SG}}^b_{\overline a}(i\pi/2-\theta)=
{S_{SG}}^{ab}_{a'b'}(2\theta){K^{(\epsilon)}_{SG}}^{a'}_{\overline
b'}(i\pi/2+\theta)\label{crossunit}\ ,\eeqa
as the ``$in$'' and ``$out$'' states are related through the
$S$-matrix. Notice that this relation is linear in the boundary
operators. For simplicity, let us introduce
 two meromorphic functions $Y_0(\theta)$ and $Y_1(\theta)$ which solve the
 functional equations ($u\equiv-i\theta$):
\beqa &&Y_0(\theta)Y_0(-\theta)=1\qquad \qquad \qquad \qquad
\qquad\qquad \qquad \mbox{and}\qquad
Y_0(i\pi/2-\theta)=\sin(\lambda(\pi-2u)) \rho(2u) Y_0(i\pi/2+\theta)\ ;\ \nonumber \\
&&Y_1(\theta)Y_1(-\theta)=\big[\sinh^2(2\lambda\theta)-\sinh^2(i\pi\lambda)\big]^{-1}\qquad
\mbox{and}\qquad Y_1(i\pi/2-\theta)=Y_1(i\pi/2+\theta)\
.\nonumber\eeqa
Then, it is not difficult  to show that for the choice
\beqa Y(\theta)=Y_0(\theta)Y_1(\theta)\label{Y}\ \qquad \eeqa
in (\ref{ksg}), the boundary unitarity and boundary
cross-unitarity conditions are satisfied. Following
Ghoshal-Zamolodchikov notations we finally obtain
\beqa Y_0(\theta)=R_0(u)G_0(u)\eeqa
where we used \cite{Gho93}
\bea R_0(u)=\prod_{k=1}^{\infty}\Big[\frac{\Gamma(4\lambda k
-2\lambda u/\pi)\Gamma(1+4\lambda(k-1) -2\lambda
u/\pi)}{\Gamma(\lambda(4k-3) -2\lambda
u/\pi)\Gamma(1+\lambda(4k-1) -2\lambda
u/\pi)}/(u\rightarrow-u)\Big]\eea
and we introduced
\beqa G_0(u)=
\prod_{k=1}^{\infty}\Big[\frac{\Gamma(1+(4k-2)\lambda-2\lambda
u/\pi)\Gamma((4k-2)\lambda-2\lambda
u/\pi)}{\Gamma((4k-4)\lambda-2\lambda
u/\pi)\Gamma(1+4k\lambda-2\lambda u/\pi)}/(u\rightarrow-u)\Big]\ .
\ \nonumber\eeqa
Notice that the factor $R_0(u)$ contains poles in the ``physical
strip''\ \  $0< u < \pi/2$\ \ located at \ $u_n=n\pi/2\lambda$\ \
for \ $n=1,...<\lambda$\ . They are associated with zero-momentum
soliton-antisoliton bound states (see last section). Also, we find
\beqa Y_1(\theta)=
\frac{\sigma(\pi/2+\pi\lambda/2,u)\sigma(\pi\lambda/2,u)}
{\sin(\pi\lambda)}\ \eeqa
where \cite{Gho93}
\bea \sigma(x,v) = \frac{\cos x}{\cos(x+\lambda
v)}\prod_{l=1}^{\infty}\Big[\frac{\Gamma(1/2+(2l-1)\lambda+x/\pi-\lambda
v/\pi)\Gamma(1/2+(2l-1)\lambda-x/\pi-\lambda
v/\pi)}{\Gamma(1/2+(2l-2)\lambda-x/\pi-\lambda
v/\pi)\Gamma(1/2+2l\lambda+x/\pi-\lambda
v/\pi)}/(v\rightarrow-v)\Big]\eea
satisfies the relations
\bea \sigma(x,v)\sigma(x,-v)= \cos^2x/(\cos(x+\lambda
v)\cos(x-\lambda v))\ , \qquad \qquad
\sigma(x,\pi/2-v)=\sigma(x,\pi/2+v)\ .\eea
 From (\ref{ksg}) it is clear that
there are no resonance states in comparison with the non-dynamical
SG model. However, the factors \ $\sigma(\pi/2+\pi\lambda/2,u)$\
 and \ $\sigma(\pi\lambda/2,u)$ \ brings an
infinite number of singularities at {\it real} values of $u$. In
the ``physical strip''\ \ $0< u <\pi/2$,\ \ depending on the value
of the SG coupling constant, they are located at:
\bea u_k^{even}=\frac{\pi}{2}-\frac{k\pi}{\lambda}\qquad
\mbox{for} \qquad 0 < k < \frac{\lambda}{2}\ \qquad \ \mbox{and}\
\qquad u_k^{odd}=\frac{\pi}{2} - \frac{(2k-1)\pi}{2\lambda} \qquad
\mbox{for} \qquad 0 < k < \frac{\lambda+1}{2}\ .\eea
Denoting $E_0$ as the ground state energy, we identify these poles
as boundary bound states with energy
$E^{_{odd}^{even}}_k-E_0=M\cos(u^{_{odd}^{even}}_k)$. For any $k$,
we conclude that the boundary bound states are stable.

\subsection{Bound state reflection matrix for the dynamical boundary}
Above, we considered the reflection matrix of the
soliton/antisoliton off the boundary. We now turn to the
reflection matrix of the bound states (breathers) $B_n(\theta)$.
To calculate it, we use the boundary bootstrap method introduced
in \cite{Gho93,Fri95} and mainly follow \cite{Gho94}. Using
``particle creation'' operator formalism, the reflection matrix
defines the commutation relation
\bea B_n(\theta)B=R^{(n)}_B B_n(-\theta)B\eea
where $B_n(\theta)$ creates the bound state $B_n$ with rapidity
$\theta$. Then,  the boundary bootstrap equation reads
\cite{Gho93}
\beqa
f^n_{i_1i_2}{K^{(\epsilon)}_{SG}}_{j_1}^{i_1}(\th+\th_n/2){S_{SG}}_{j_2f_1}^{i_2j_1}(2\theta)
{K^{(\epsilon)}_{SG}}_{f_2}^{j_2}(\th-\th_n/2)=f^n_{f_1f_2}R^{(n)}_B(\theta)\label{boot}\eeqa
where $f^n_{i_1i_2}$ are vertices satisfying \
$f_{+-}^n(-1)^n=f_{-+}^n$\  and \ $f^n_{\pm\pm}=0$. Furthermore,
the solution of (\ref{boot}) has to satisfy the boundary unitarity
and boundary cross-unitarity conditions
\beqa &&R^{(n)}_B(\theta)R^{(n)}_B(-\theta)=\mathbb{I}\
,\nonumber\\
&&R^{(n)}_B(i\pi/2-\theta)=R^{(n)}_B(i\pi/2+\theta)S^{(n,n)}(2\theta)\
, \eeqa
respectively. Here, $S^{(n,n)}(2\theta)$ denotes the scattering
amplitude for $B_n+B_n\rightarrow B_n+B_n$ process \cite{Zam79}.
In particular, it possesses a pole in the physical strip located
at $\theta=n\pi/2\lambda$. It is straightforward to show that
\bea &&Y_0(\theta + \theta_n/2) Y_0(\theta - \theta_n/2)
\rho(-2i\theta)/i= -\frac{1}{\sinh(2\lambda
\theta-i\lambda\pi)}R_0^{(n)}(u)S^{(n)}(0,u)S^{(n)}(\pi/2,u)\
,\nonumber \\
&&Y_1(\theta + \theta_n/2) Y_1(\theta - \theta_n/2)=
\frac{1}{\sinh(2\lambda\theta)\sinh(2\lambda\theta-2i\lambda\pi)}
S^{(n)}(\pi/2+\pi\lambda/2,u)S^{(n)}(\pi\lambda/2,u)\ .\nonumber
\eea
Here we used the notations of \cite{Dor00,Baj02} i.e.
\bea
R_0^{(n)}(u)=\frac{\big(\frac{1}{2}\big)\big(1+\frac{n}{2\lambda}\big)}
{\big(\frac{3}{2}+\frac{n}{2\lambda}\big)}\prod_{l=1}^{n-1}\frac{\big(\frac{l}{2\lambda}\big)\big(1+\frac{l}{2\lambda}\big)}
{\big(\frac{3}{2}+\frac{l}{2\lambda}\big)^2}\qquad \mbox{and}
\qquad
S^{(n)}(x,u)=\prod_{l=0}^{n-1}\frac{\big(\frac{x}{\lambda\pi}-\frac{1}{2}+\frac{n-2l-1}{2\lambda}\big)}
{\big(\frac{x}{\lambda\pi}+\frac{1}{2}+\frac{n-2l-1}{2\lambda}\big)}
\eea
where we have the shorthand notation \ $\big(x\big)=
\sin(u/2+x\pi/2)/\sin(u/2-x\pi/2)$. Taking into account the
non-diagonal part of the soliton reflection matrix, we finally
obtain after some calculations the result
\beqa R_{B}^{(n)}(\theta)=(-1)^n R_0^{(n)}(u)
S^{(n)}(0,u)S^{(n)}(\pi/2,u)S^{(n)}(\pi\lambda/2+\pi/2,u)
S^{(n)}(\pi\lambda/2,u)\ \label{RnB}\eeqa
which, clearly, does not depend on the boundary operators
$\hE_{\pm}$. As expected, the factor $R_0^{(n)}(u)$ contains the
poles in the physical strip located at $u_n=\pi/2-n\pi/2\lambda$
for $\lambda>1$. They are associated with the breather-breather
bound states $B_{2n}$.

\subsection{Restriction of the boundary state}
Up to now, we only considered the scattering properties without
imposing any restriction beyond (\ref{cond1}) to the boundary
state $|B\rangle$. However, it is necessary to exclude those which
would violate certain physical principles, i.e. it must be in
accordance with the SG scattering processes. For instance, the
term $R_0(-i\theta)$ in eq. (\ref{Y}) contains poles in the
``physical strip'' \  $0< \theta < i \pi/2$ \ located at
$\theta_n=in\pi/2\lambda$; $n=1,...<\lambda$. They are associated
with the zero-momentum breathers $B_n(0)$ that should appear as
bound states of soliton-antisoliton. At these values of the
rapidity any off-diagonal entry of (\ref{ksg}) which appears in
(\ref{stategen}) applied to a state $|{\small\textsf{B}}_n\rangle$
will not, in general, give a null result. As breathers are
neutral, the state $|{\small\textsf{B}}_n\rangle$ must however be
a null vector of $B(\theta_n)$ and $D(\theta_n)$, i.e it must obey
the condition
\beqa B(\theta_n)|{\small\textsf{B}}_n\rangle=0\ \ \quad \ \
\mbox{and}\quad\ \ \ \
D(\theta_n)|{\small\textsf{B}}_n\rangle=0\eeqa
for all $n=1,2,...<\lambda$. In the non-dynamical boundary case
\cite{Gho93} this was trivially satisfied due to the form of the
off-diagonal entries. Combining these two conditions we obtain
\beqa \big[\cosh\pt,\cosh\qt\big]|{\small\textsf{B}}_n\rangle=0
\qquad \mbox{and}\qquad \cosh\pt\
\cosh\qt|{\small\textsf{B}}_n\rangle =
(-1)^{n}\cos^2(\pi\lambda)|{\small\textsf{B}}_n\rangle\label{cond}\
.\eeqa
To characterize each state $|{\small\textsf{B}}_n\rangle$
explicitly, we introduce the one dimensional ``space''
representation $\{|\textsf{x}\rangle\}$ for the Heisenberg
operators defined by
\beqa \pt|\textsf{x}\rangle=-2i\pi\lambda\partial/\partial
\textsf{x} \ ,\qquad
\qt|\textsf{x}\rangle=\textsf{x}|\textsf{x}\rangle\qquad
\mbox{and} \qquad
\langle\textsf{x}|{\small\textsf{B}}_n\rangle\equiv
\Psi_n(-i\textsf{x}/\pi)\ .\eeqa
From the first equation in (\ref{cond}), it is straightforward to
obtain a difference equation satisfied by $\Psi_n(\textsf{y})$ for
$\textsf{y}=-i\textsf{x}/\pi$. Obviously, at that stage $n$ does
not appear explicitly. It reads
\beqa
\frac{\Psi_n(\textsf{y}+2\lambda)}{\Psi_n(\textsf{y}-2\lambda)}=\frac{\sin(\pi(\textsf{y}-\lambda))}
{\sin(\pi(\textsf{y}+\lambda))}\label{diffeq0}\ . \eeqa
Up to an irrelevant overall constant, it can be solved by the
integral representation
\beqa
\Psi_n(\textsf{y})=\exp\int_{0}^{\infty}\frac{dt}{t}\frac{\sinh^2((\textsf{y}-1/2)t)}{\sinh(t)\cosh(2\lambda
t)}\qquad \qquad \mbox{for}\qquad {\cal
R}e(\textsf{y})<1+\lambda\label{psin}\eeqa
and is defined through analytic continuation otherwise. The second
equation will fix the ``allowed'' values of $\textsf{y}$. It reads
\beqa (-1)^n
2\cos^2(\pi\lambda)\Psi_n(\textsf{y})=\cos(\pi\textsf{y})\big(\Psi_n(\textsf{y}-2\lambda)+
\Psi_n(\textsf{y}+2\lambda)\big)\qquad \mbox{for}\qquad
\textsf{y}\in \{\textsf{y}^{(n)}\}\ .\label{diffeq}\eeqa
The condition (\ref{diffeq}) is divided into two sectors, ``odd''
$n$ breathers and ``even'' $n$ breathers. Then, using the
definition (\ref{psin}) the boundary state
$|{\small\textsf{B}}_n\rangle$ can be decomposed on the finite set
of elementary ``odd'' and ``even'' states $n\in\{2p,2p+1\}$ \
with\ \ $|{\small\textsf{B}}_n\rangle
={\small\textsf{N}_n}\Psi_n(\textsf{y}^{(n)})|\textsf{y}^{(n)}\rangle$
\ \ (${\small\textsf{N}}_n$ are some normalization constants)
which, due to (\ref{stategen}), implies that\ \
$K^{(\epsilon)}_{SG}(\theta_n)|{\small\textsf{B}}_n\rangle$\ \
will be automatically block diagonal.

Notice that in the commutative limit $\lambda=0$ in
(\ref{diffeq0}) the function $\Psi_n(y)$ remains arbitrary and
(\ref{diffeq}) leads to $y=n$. In other words, the boundary state
$|B\rangle$ is not restricted anymore in agreement with
\cite{Gho93}.

\section{Concluding remarks}
As was suggested in \cite{Gho93}, boundary integrable quantum
field theories including degrees of freedom at the boundary can be
constructed. Here, we provide such an example corresponding to a
sine-Gordon quantum field theory coupled to a non-linear quantum
oscillator at the boundary, which follows the work initiated in
\cite{Bas01}. For certain boundary conditions, integrability of
the model is preserved at quantum level for an exact relation
between the sine-Gordon coupling constant and the boundary
quantization length. In particular, this later quantity fixes the
overall sign of the boundary perturbation. The corresponding
soliton/antisoliton reflection scattering matrix
$K^{(\pm)}_{SG}(\theta)$ is constructed explicitly using an
extension to the dynamical boundary case of the method based on
boundary non-local conserved charges \cite{Mez98,Del02}. Bound
state reflection matrices are also constructed explicitly. Using
some physical constraints, the general form of the boundary state
is shown to be restricted.

To study this model further, extensions to the dynamical case of
known nonperturbative approaches could be useful. In the
non-dynamical case, they have provided an efficient way of
studying boundary effects in finite-size system. For instance, TBA
analysis \cite{Lec95} or boundary reflection amplitude method
\cite{ahn} from which the effective central charge and boundary
ground state energy can be deduced. Comparison of both approaches
provides important checks of the exact relation between UV and IR
parameters as well as the boundary ground state energy. Here, such
methods extended to the dynamical case would obviously provide
important information.\vspace{0.2cm}

Due to the form of our model, one may wonder if it could not be
related with certain massive extension of the field theory
description of the Kondo model \cite{Fen96}. In \cite{Bass99} two
massive versions of the anisotropic spin $1/2$ Kondo model have
been proposed. It is thus interesting to compare results from
\cite{Bass99} (see also \cite{Lec99}) to ours. The first massive
Kondo (MK) model is one natural generalization of the integrable
massless Kondo model for arbitrary spin studied in \cite{Fen96}.
The MK Hamiltonian reads
\beqa H_{MK}=\frac{1}{2}\int_{-\infty}^0 dx \Big((\pi(x))^2
+(\partial_x\phi(x))^2- G\cos(\beta\phi(x))\Big) + \lambda_0
\big(S_+e^{i\beta\phi(0)/2} + S_-e^{-i\beta\phi(0)/2}\big) \ \eeqa
where $S_{\pm}$ form a spin $1/2$ representation of the
$q_0-$deformed quantum algebra $su_{q_0}(2)$. The $su_{q_0}(2)$
relation are
\beqa [S_z,S_\pm]=\pm 2S_\pm \ , \qquad \qquad
[S_+,S_-]=\frac{q_0^{S_z}-q_0^{-S_z}}{q_0-q_0^{-1}}\qquad \qquad
\mbox{and}\qquad \qquad q_0=\exp(i\beta^2/8)\ . \eeqa
This model has been studied in details in \cite{Lec99} at the free
fermion point $\beta^2=4\pi$ and in \cite{Bass99} at the
reflectionless points for which it was claimed to be integrable.
Unlike the massless version, it is not integrable for arbitrary
$\beta$. On the other hand, in \cite{Bass99} the modified massive
Kondo model (mMK) with Hamiltonian
\beqa H^{mMK}&=&\frac{1}{2}\int_{-\infty}^0 dx \Big((\pi(x))^2
+(\partial_x\phi(x))^2- G\cos(\beta\phi(x))\big)\nonumber \\ &&
\qquad \qquad \qquad \qquad -i\lambda_0
\big(S_+\cos(\beta(\phi(0)-\hphi)/2) -
S_-\cos(\beta(\phi(0)-\hphi^*)/2)\Big) \ \eeqa
was proposed. Here
$\hphi=-\frac{2}{\beta}(\frac{\pi}{2}-i\hphi_0)$ with $\hphi_0$ a
free parameter and the coupling $\lambda_0$ is related to the bulk
coupling $G$. This model is believed to be integrable for
arbitrary $\beta$ and a reflection matrix has been conjectured.

 As was mentioned in \cite{Lec99,Bass99}, it is important to notice
that the analysis supposed that the boundary scattering does not
involve boundary operators. Here we didn't assume this condition,
which is crucial for our analysis and leads to a new solution
(\ref{ksg}) of the boundary Yang-Baxter equations. Then, it would
be interesting to see in such case if more solutions can be
obtained if a ``simpler'' bulk scattering is considered instead
(reflectionless points). Notice that the free fermion point and
the reflectionless points are special cases for which our
operators (\ref{defope0}) $\hE_\pm$ commute, i.e. the $K_0$
reflection matrix reduces to Ghoshal-Zamolodchikov and
DeVega-Gonzales-Ruiz \cite{Gho93,DeV94} result (see end of section
3).

Other remarks can be made. First, the model mMK is integrable for
arbitrary $\beta$ only for a special relation between $G$ and
$\lambda_0$, and the deformation parameter $q_0$ is expressed in
terms of the SG coupling constant $\beta$. It is also the case for
our model. Secondly, in the MK and mMK models the boundary
dependence only occurs in the CDD factors. With the boundary
dependence isolated, the massless limit follows by taking the
limit $\theta\rightarrow\infty$, which leads to the (massless)
Kondo model reflection matrix. However, in our model the boundary
dependence occurs through the operators $\cosh(\pt)$, $\cosh(\qt)$
in the minimal (non-CDD part) of the reflection matrix
(\ref{ksg}). Consequently, taking the massless limit of our model
requires an analysis which goes beyond the scope of this paper.
Finally, for $q$ a root of unity it is easy to construct a
finite-dimensional cyclic representation for $\hE_{\pm}$. This
property might have a certain interest: partition functions of the
massless boundary sine-Gordon and the anisotropic Kondo model can
be related if the boundary spin is in a cyclic representation
\cite{Fen96}.\vspace{0.2cm}

Among other models, it is now quite natural to consider the
analytic continuation of the boundary sine-Gordon model studied
above. It should provide one of the simplest boundary integrable
QFT, the sinh-Gordon model restricted to $x<0$ with quantum
degrees of freedom at the boundary. Its spectrum consists of only
one particle with mass $m$ and the bulk scattering $S$-matrix for
a pair of particles reads \cite{Ari79}
\bea
S_{shG}(\theta)=\frac{\big(-1\big)}{\big(\frac{B}{2}\big)\big(1-\frac{B}{2}\big)}\qquad
\mbox{where}\qquad B=\frac{2b^2}{2+b^2} \ .\eea
In particular, this amplitude is self-dual under the weak-strong
coupling duality transformation $b\leftrightarrow 2/b$. Notice
that the bulk $S$-matrices associated with higher simply laced
Toda theories enjoy the same property. As was shown in
\cite{Cor98,Cor99,Cor00} for non-dynamical integrable boundary
conditions, the sinh-Gordon reflection factor obtained from
analytic continuation of Ghoshal's result \cite{Gho94} is not
self-dual.

Here, taking the analytic continuation $\betah\rightarrow ib$ in
(\ref{RnB}) for $n=1$, the fundamental particle reflection factor
takes the simple form
\bea
K_{shG}(\theta)=\frac{\big(-\frac{1}{2}\big)}{\big(\frac{B}{4}\big)\big(\frac{1}{2}-\frac{B}{4}\big)}\
. \eea
Then, up to an overall sign in front of the boundary perturbation,
integrability is ensured for the boundary quantization length
(after changing $\betah\rightarrow ib$ in (\ref{relat}) $(-)$)
fixed to
\beqa \big[\pt,\qt\big]=\alpha_{shG}\qquad \ \ \ \ \mbox{with}\ \
\ \ \qquad \alpha_{shG}=i4\pi/B\ .\eeqa
Consequently, weak-strong coupling duality of the boundary
integrable sinh-Gordon model corresponds to the boundary
operator-coupling transformation
\beqa \{\pt^*,\qt^*,b^*\}=\{b\pt/\sqrt{2},b\qt/\sqrt{2},2/b\}\
.\eeqa

Although we didn't discuss them here, it seems to us that our
analysis can provide a starting point concerning certain open
problems:\vspace{0.2cm}\\
$\bullet$ A different choice of trigonometric $R$-matrix gives the
Boltzmann weights of the six-vertex model. It is known to be
related to the XXZ and the XXX (in its rational limit) spin
chains. Applying  our result to this case,  the value of the
boundary quantization length will define the value of the
anisotropy which preserves integrability of the model.
Consequently, we hope our solution $K_0(\theta)$ which satisfies
the corresponding qBYBEs can be useful in the study of spin chains
with dynamical boundary conditions\,\footnote{We thank A. Doikou for bringing our attention to \cite{nepo,doi}.}.\\
$\bullet$ Statistical models with extended line of defects have
attracted attention as such inhomogeneities affect the critical
properties of the pure statistical systems. In the continuum
limit, the scattering theory of their massive excitations is
described in terms of the bulk scattering amplitudes as well as
those associated with the interaction of particles with the defect
line. If integrability is preserved, they reduce to
reflection-transmission amplitudes which satisfy the so-called
reflection-transmission equations \cite{muss}. Similarly to
\cite{Bas01} and in the present work, one may be interested in
solutions of these equations with quantum boundary degrees of
freedom in the defect, taking a reflection matrix of the form
(\ref{def}), (\ref{entrees}). We will discuss this problem
elsewhere
\cite{prep}.\\

\noindent{\bf Acknowledgements:} We are very grateful to R. Sasaki
for a careful reading of the manuscript and important comments.
P.B also thanks P. Dorey, A. Doikou and M. Stanishkov for useful
discussions and interest in this work, as well as S. Nicolis and
S. Jhingan. P.B thanks hospitality of the ``Laboratoire de
Math\'ematiques et Physique Th\'eorique'' in Tours where part of
this work has been done. K.K thanks hospitality of Yukawa
Institute for Theoretical Physics. P.B's work supported by JSPS
fellowship.


\begin{thebibliography}{10}
%
\bibitem{Skl87} E.K. Sklyanin, Funct. Anal. Appl. {\bf 21} (1987) 164.
%
\bibitem{Skl88} E.K. Sklyanin, J. Phys. {\bf A 21} (1988) 2375.
%
\bibitem{Gho93} S. Ghoshal and A.B. Zamolodchikov, Int. J. Mod. Phys. {\bf A 9} (1994) {3841}.
%
\bibitem{Gho94} S. Ghoshal, Int. J. Mod. Phys. {\bf A 9} (1994) 4801.
%
\bibitem{Fri95}
A. Fring and R. Koberle, Int. J. Mod. Phys. {\bf A 10} (1995) 739.
%
\bibitem{Fen94}
 P. Fendley, H. Saleur and N.P. Warner, Nucl. Phys. {\bf B 430} (1994) 577.
%
\bibitem{Sal95}
 H. Saleur, S. Skorik and N.P. Warner, Nucl. Phys. {\bf B 441} (1995) 421.
%
\bibitem{Sko95}
 S. Skorik and H. Saleur, J. Phys. {\bf A 28} (1995) 6605.
%
\bibitem{Lec95}
A. LeClair, G. Mussardo, H. Saleur and S. Skorik, Nucl. Phys. {\bf
B 453} (1995) 581.
%
\bibitem{Fat97}
 V.A. Fateev, S. Lukyanov, A.B. Zamolodchikov and Al.B.
 Zamolodchikov, Phys. Lett. {\bf B 406} (1997) 83.
%
\bibitem{Dor99}
 P. Dorey, R. Tateo and G. Watts, Phys. Lett. {\bf B 448} (1999)
 249.
%
\bibitem{Cor98}
E. Corrigan, Int. J. Mod. Phys. {\bf A 13} (1998) 2709.
%
\bibitem{Cor99}
 E. Corrigan and G.W. Delius, J. Phys. {\bf A 32} (1999) 8601.
%
\bibitem{Cor00} E. Corrigan and A. Taormina, J. Phys. {\bf A 33} (2000) 8739.
%
\bibitem{Dor00} P. Mattsson and P. Dorey, J. Phys. {\bf A 33} (2000) 9065.
%
\bibitem{Baj02} Z. Bajnok, L. Palla, G. Takacs and G.Zs. Toth,  Nucl. Phys.
{\bf B 622} (2002) 548.
%
\bibitem{Mat} Peter Mattsson, Ph.D. thesis, Durham.
%
\bibitem{Bas01}
P. Baseilhac and G.W. Delius, J. Phys. {\bf A 34} (2001) 8259.
%
\bibitem{Kuz95}
V.B. Kuznetsov, J. Phys. {\bf A 28} (1995) 4639.
%
\bibitem{Kuz96}
V.B. Kuznetsov and A.V. Tsyganov, J. Math. Sci. {\bf 80} (1996)
1802.
%
\bibitem{Ber91}
D. Bernard and A. LeClair, Commun. Math. Phys. {\bf 142} (1991)
99.
%
\bibitem{Mez98}
L. Mezincescu and R.I. Nepomechie, Int. J. Mod. Phys. {\bf A 13}
(1998) 2747.
%
\bibitem{Del02}
G.W. Delius and N. MacKay, ``Quantum group symmetry in sine-Gordon
and affine Toda field theories on the half-line'', submitted to
Commun. Math. Phys., hep-th/0112023.
%
\bibitem{DeV94}
 H.J. de Vega and A. Gonz\'alez-Ruiz, J. Phys. {\bf A 27} (1994) 6129.
%
\bibitem{Hol92}
T. Hollowood, Nucl. Phys. {\bf B 384} (1992) 523.
%
\bibitem{Zam95}
Al.B. Zamolodchikov, Int. J. Mod. Phys. {\bf A 10} (1995) 1125.
%
\bibitem{Zam79}
A.B. Zamolodchikov and Al.B. Zamolodchikov, Ann. Phys. {\bf 120}
(1979) 253.
%
\bibitem{ahn} C. Ahn, C. Kim and C. Rim, Nucl. Phys. {\bf B 628} (2002)
486.
%
\bibitem{Fen96}
P. Fendley, F. Lesage and H. Saleur, J. Stat. Phys. {\bf 85}
(1996) 211;\\
P. Fendley and H. Saleur, Phys. Rev. Lett. {\bf 75} (1995) 4492.
%
\bibitem{Bass99}
Z.S. Bassi and A. LeClair, Nucl.Phys. {\bf B 552} (1999) 643.
%
\bibitem{Lec99}
A. LeClair, Ann. Phys. {\bf 271} (1999) 268.
%
\bibitem{Ari79}
A.E. Arinshtein, V.A. Fateev and A.B. Zamolodchikov, Phys. Lett.
{\bf B 87} (1979) 389.
%
\bibitem{nepo}
R.I. Nepomechie, Nucl. Phys. {\bf B 622} (2002) 615.
%
\bibitem{doi}
A. Doikou and P.P. Martin, ``Hecke algebraic approach to the
reflection equation for spin chains'', hep-th/0206076.
%
\bibitem{muss}
G. Delfino, G. Mussardo and P. Simonetti, Phys. Lett. {\bf B 328}
(1994) 123;\\
%
G. Delfino, G. Mussardo and P. Simonetti, Nucl. Phys. {\bf B 432}
(1994) 518.
%
\bibitem{prep}
P. Baseilhac and K. Koizumi, in preparation.














\end{thebibliography}
\end{document}